\documentclass{PoS}

\usepackage[round]{natbib}
\usepackage{url}

\newcommand{\psr}{PSR~J1745--2900}

\newcommand{\sgra}{Sgr~A$^\star$}
\newcommand\farcs{\hbox{$.\!\!^{\prime\prime}$}}
\newcommand{\onefig}[1]{\includegraphics[width=0.2\textwidth,trim={1cm
      0cm 1.3cm 0cm},clip]{#1}}

\title{Probing interstellar scattering towards the Galactic centre
  with pulsar VLBI\thanks{Original title of the conference talk:
    \emph{Scattering as a nuisance and as a tool}}}
\ShortTitle{Interstellar scattering towards the Galactic centre}

\author{\speaker{Olaf Wucknitz}\\ Max-Planck-Institut f\"ur
  Radioastronomie, Auf dem H\"ugel 69, 53121 Bonn, Germany,\\ E-mail:
  \email{wucknitz@mpifr-bonn.mpg.de}}

\abstract{Temporal scatter-broadening can seriously affect our ability
  to find pulsars orbiting the central mass in our Galaxy. Many of
  these invaluable probes of geometry around the black hole are
  expected, but none have been found in close orbits so far, possibly
  as result of strong scattering. The magnetar \psr\ discovered in
  2013 at a separation of < 3 arcsec is not the optimal type of pulsar
  for studies of general relativity, but it can be used to investigate
  the scattering properties so that search strategies can be adapted
  accordingly.

  This contribution presents an observation of this magnetar using
  short baselines between VLBI stations in Europe in a non-standard
  interferometry mode. The most important goal is determining the
  distance of the scattering screen, or the distribution of scattering
  material if not confined to one screen.

The analysis is based on phase-binned visibilities that allow
measuring the shape of the scattering disk and how it grows with
increasing delay over the scattering tail of the pulse
profile. Narrow rings growing with the square root of delay are
expected for a single thin scattering screen and the preliminary
results are indeed consistent with this expectation. This means that
most of the angular and temporal broadening is caused by the same and
relatively thin scattering screen and that, in contrast to standard
models of the interstellar scattering behaviour near the Galactic
centre, this screen is located about halfway between the centre and
us.}

\FullConference{12th European VLBI Network Symposium and Users Meeting,\\
		7-10 October 2014\\
		Cagliari, Italy}

\begin{document}

\section{Pulsars near black holes}

The motivation for this project is the fact that pulsars closely
orbiting a black hole can potentially be used to test general
relativity and to determine parameters of the black hole with the
highest precision. For these tests, we want to probe the geometry in
strong gravitational fields (thus near a black hole). Our most
accurate measurements are those of time, and in astronomy these also
happen to be comparatively simple. It is thus a natural idea to use
highly accurate clocks as probes of geometry and measure their `ticks'
for the analysis. Luckily nature is so kind to provide such accurate
clocks in the form of pulsars.

One of the most promising candidate for a black hole within reasonable
distance is the one in \sgra\ in the Galactic centre (GC) at a
distance of about 8.5\,kpc with a mass around 4 million Solar
masses. The density of stars in that region is very high, and common
formation scenarios predict an abundance of neutron stars and in
particular pulsars in orbit around \sgra.

\citet{liu12} and others describe in detail, which parameters can be
determined, e.g.\ the mass and spin of the GC black hole, allowing us to
test the cosmic censorship conjecture, or the quadrupole moment so
that the no-hair-theorem can be tested. Additionally perturbations
from other masses can also be detected easily.

A potential major difficulty in finding these pulsars is the effect of
interstellar scattering producing subimages that are deflected and
thus delayed by different amounts. Such a range of delays can easily
wash out the `pulsations' that are used to identify pulsars, that form
their clock ticks and are thus the basis of accurate timing
measurements. Unfortunately the GC appears to be the region affected
the worst, so that observing parameters have to be chosen that minimise
the effects.

The deflection angles roughly scale with the observing wavelength as
$\theta\propto\lambda^2$ and the delays with
$\tau\propto\theta^2\propto\lambda^4$, so that the effect becomes much
stronger at low frequencies and that higher-frequency observations may
be needed.\footnote{For Kolmogorov-type turbulence, the exponents are
  even slightly higher (2.2 and 4.4).} Exact numbers for the GC are
highly model-dependent, but were thought to be as bad as hundreds of
seconds for frequencies around 1\,GHz, which would wash out the
signals even from the slowest pulsars \citep{cordes97}.

It has been tried to search for pulsars near the GC at high
frequencies, e.g.\ with the GBT at 15\,GHz \citep{macquart10} and with
Effelsberg at 19\,GHz \citep{eatough13a}. Despite the
expectations that at least dozens of pulsars should have been found,
no single object could be detected in these surveys within 1--2\,pc of
the GC.

\section{Interstellar scattering towards the Galactic centre}

\begin{figure}[hbt]
\centering\includegraphics[width=0.5\textwidth]{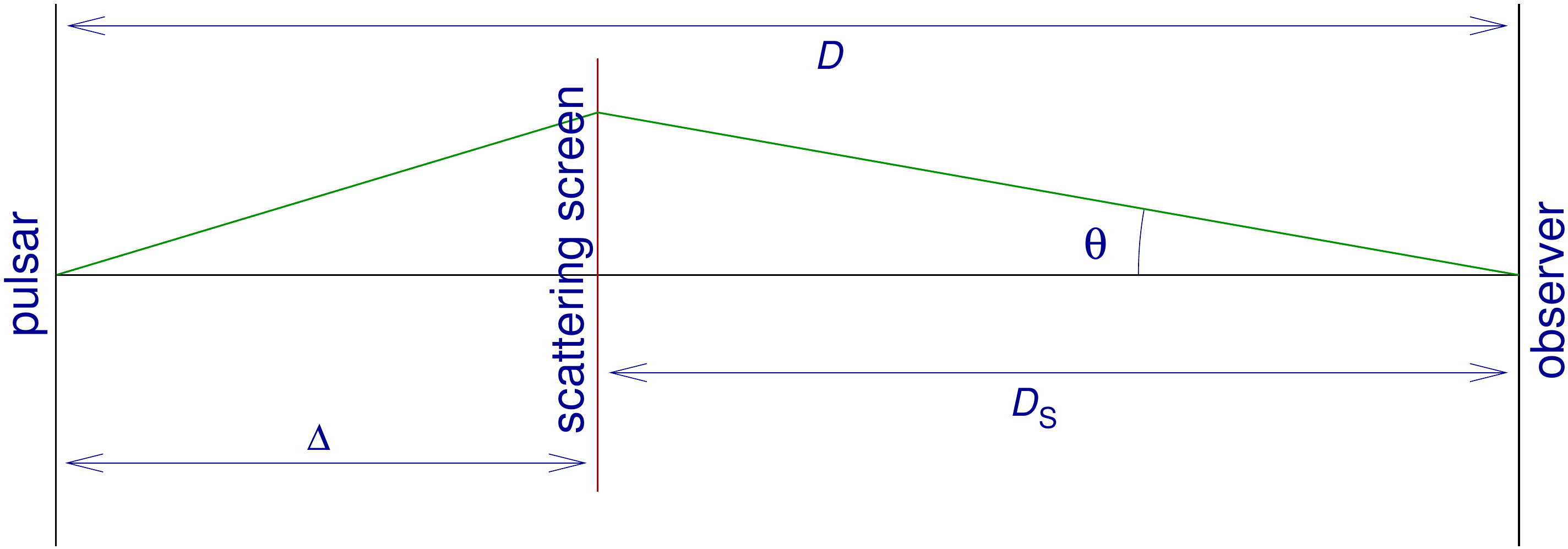}
\caption{The fundamental scattering geometry with one subimage deflected by an angle $\theta$, which causes a geometric delay of $\tau$.}
\label{fig:geometry}
\end{figure}

From the scattering geometry (Fig.~\ref{fig:geometry}) the relation between scattering angle $\theta$ (leading to the size of the scattering disk for the ensemble of all subimages) and the corresponding delay $\tau$ (leading to the temporal scattering tail) can be derived easily in the small-angle approximation:
\begin{displaymath}
c\tau = \frac12 \theta^2 D'
\hspace{8em}
D' =\frac{D\,(D-\Delta)}{\Delta}
\end{displaymath}
Note how strongly the scattering delay for a given angular size
$\theta$ depends on the distance $\Delta$ to the scattering screen. If
it is very close to the pulsar (as expected for the GC), the delay
becomes very large, while it vanishes in the limit of a screen very
close to the observer. The angular scattering size of \sgra\ is about
1\,arcsec for a frequency of 1\,GHz.

Taking the model value $\Delta=( 133 ^{+200}_{~-80} ) \,$pc from
\citet{lazio98}, we would expect 150\,sec at 1\,GHz, while this
scattering time reduces to 2\,sec if the screen is halfway between the
GC and us. Knowing the location of the scattering screen (or the
distribution of scattering material) is thus of uttermost importance,
not only to understand why we have not found any pulsars in that
region until recently, but also to define our search strategy for the
future.

\section{The GC magnetar PSR J1745--2900}

In April 2013, the Swift satellite detected an X-ray flare in the area
of \sgra\ \citep{kennea13}. NuSTAR found pulsations with
a period of 3.76\,sec in the same object and identified it as a
possible magnetar \citep{mori13}. Within days, the magnetar was also
detected at radio frequencies \citep{eatough13b} with the Effelsberg
telescope and studied extensively since then \citep[e.g.][]{shannon13}.

\citet{spitler14} showed that the scattering time is proportional to
$\lambda^{3.8}$, close to the expectation, and that at 1\,GHz it is
$\tau=1.3\,$sec, much less than derived from the models of
\citet{lazio98}. At the same time \citet{bower14} measured the size of the
scattering disk with VLBI and found it to be consistent with that of
\sgra\ itself.  Combining these two results led to an estimated
distance of the scattering screen from the magnetar of
$\Delta=(5.9\pm0.3)\,$kpc, provided that the temporal and angular
scattering are caused by one and the same thin scattering
screen.

\section{Testing the one-screen hypothesis}

Because of the conflict of the results of \citet{spitler14} and
\citet{bower14} with standard GC models, the assumption of only one
scattering screen has to be tested. In an alternative scenario there
may be at least two scattering screens, one close to the GC,
dominating the temporal scattering, and one closer to us, which then
dominates the angular broadening.  By comparing only the \emph{mean}
scattering angle (angular broadening) with the \emph{mean} scattering
delay (temporal broadening), the two scenarios cannot be
distinguished. Our project aims at comparing $\theta$ with $\tau$ not
on-average but for individual parts of the scattered signal, in the
same way as described in more detail by \citet{wucknitz12} for other
targets. If both types of scattering are caused (or at least
dominated) by the same scattering screen, we expect to see the
scaling with $\tau\propto\theta^2$ if we can resolve the signal
temporally and spatially at the same time. If, on the other hand,
both effects are resulting from different screens, the temporal and
angular effects would not be correlated.

\section{Observations, correlation and calibration}

The magnetar \psr\ was observed by the LEAP\footnote{Large
  European Array for Pulsars, see \citet{ferdman10} and
  \url{http://www.leap.eu.org/}.} team on 9th Nov 2013 for about one
hour with the Effelsberg 100\,m dish, the Nan\c{c}ay radio telescope, the
phased Westerbork array and the Lovell telescope, in the frequency
range 1604--1732\,MHz using baseband pulsar backends sampling with
8~bits/sample. The analysis presented here only uses the shortest
baseline between Effelsberg and Westerbork, ranging (in projection)
from 42\,km at the beginning to 79\,km towards the end of the
scan. This is about the optimal range for the expected angular size.

Data were shipped to Bonn to be correlated using own software. In
order to determine an unknown delay offset in the Effelsberg system,
we cross-correlated the intensity variations from both stations and
used the maximum at 409\,msec as first estimate for the delay. The
data were correlated in all four polarisation products after
converting the Westerbork data to a circular basis.
The delay was then refined by fringe-fitting the correlated
visibilities. In this process and for the calibration, \sgra\ at a
distance of 2\farcs4 was used as in-beam calibrator. The calibrator
and the target were separated with binned and weighted integrations of
the correlations. For \sgra, only bins in the off-state of the pulsar
are integrated, while for \psr\ we integrate around the region of
interest (pulse peak for the calibration) and subtract as much of the
off-state to remove \sgra\ as good as possible.

We first calibrated on \sgra, taking into account delay, delay rate,
varying parallactic angle and orientation of the receivers. We also tested
the effects of dispersive delays and differential Faraday rotation due
to the inonosphere, but it turned out that explicit corrections for
these were not necessary. Bandpass correction was applied in amplitude
and phase and the fringe-fitting and calibration procedure was
re-iterated after this step. The position of the target relative to
\sgra\ was then determined by a model-fit in $uv$-space, and the
phase centre was shifted to that position.

\section{Analysis and results}

The calibrated visibilities for \psr\ were then split into bins in
pulse phase (for this preliminary analysis 0.1\,sec in width) to
represent the scattering delay (assuming an intrinsically narrow
peak), and in time and thus baseline length to represent the angular
structure.

\begin{figure}[hbt]
\centering\includegraphics[width=0.65\textwidth]{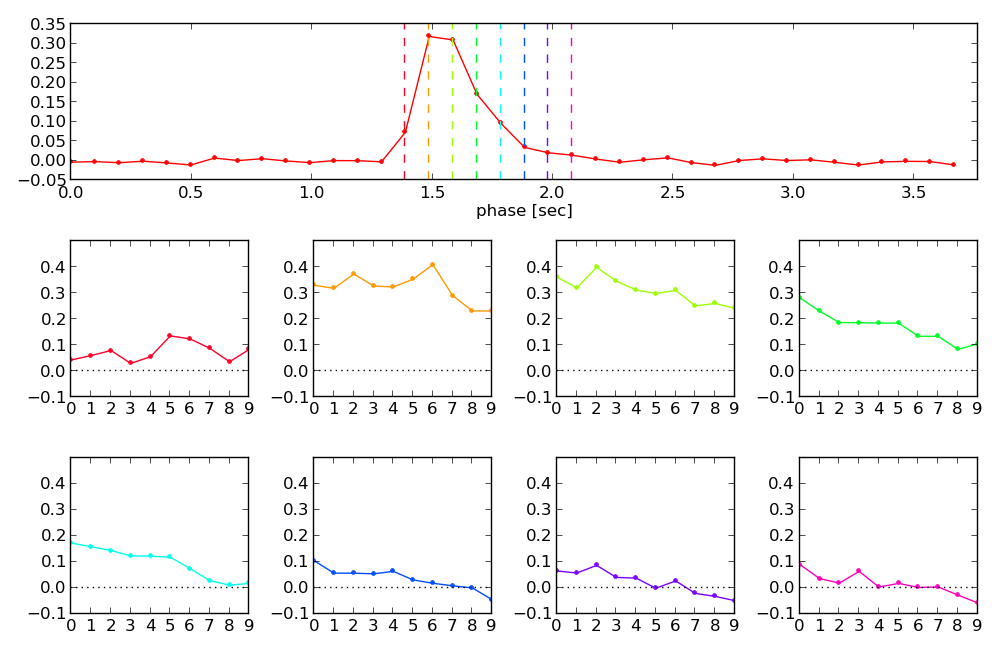}
\caption{Visibility functions for different bins of the pulse peak
  (colour-coded and labelled on top). The horizontal axes on the lower
  plots show time within the scan, over which the baseline increases
  from 42 to 79\,km. Near the peak these curves are relatively flat,
  corresponding to a compact source, but in the scattering tail they
  become steeper and even cross zero for large delays and the longest
  baselines.}
\label{fig:visib tau}
\end{figure}

\begin{figure}[hbt]
\centering%
\onefig{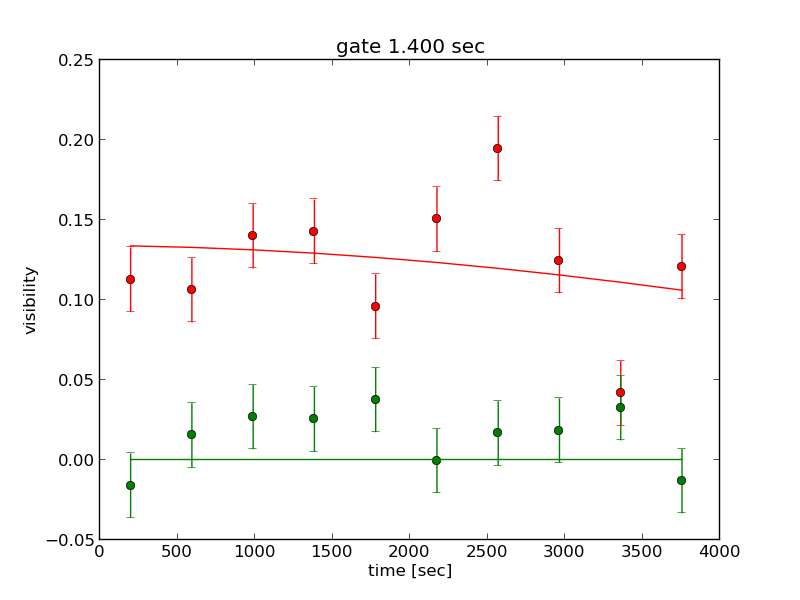}\hspace{0.5em}\onefig{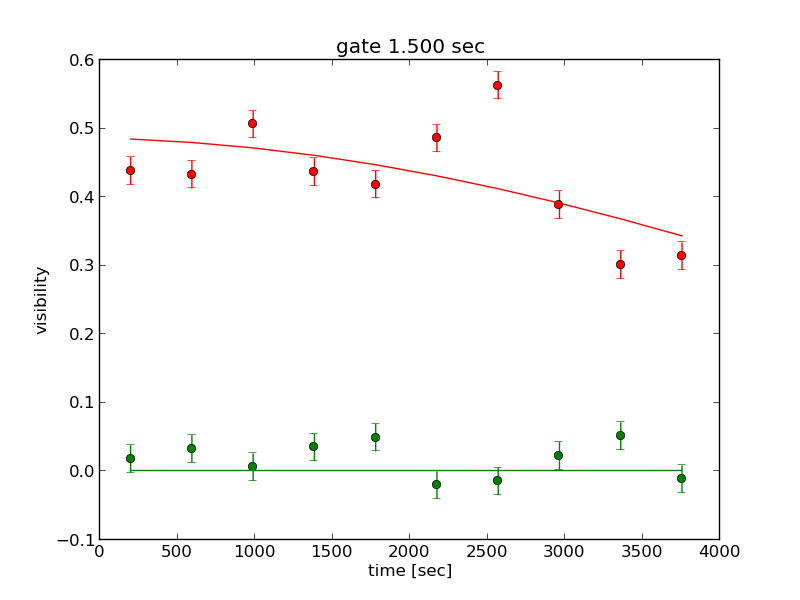}\hspace{0.5em}\onefig{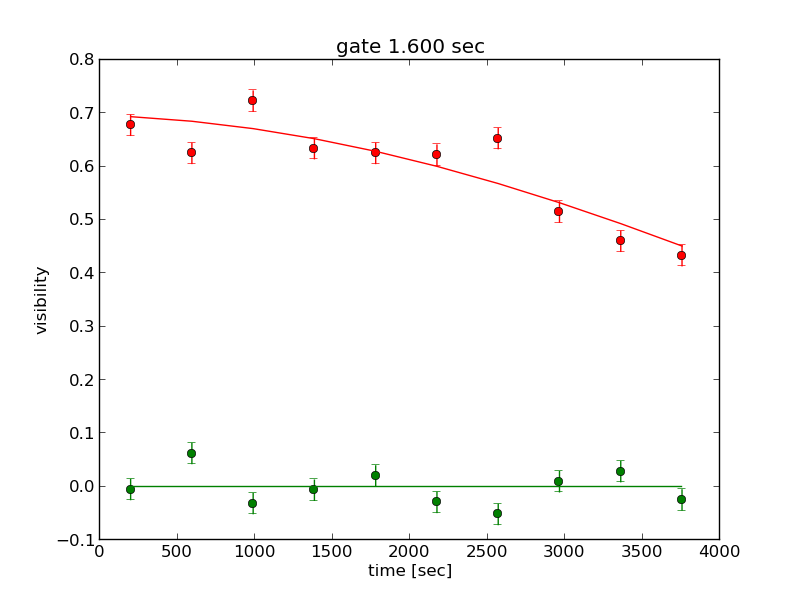}\hspace{0.5em}\onefig{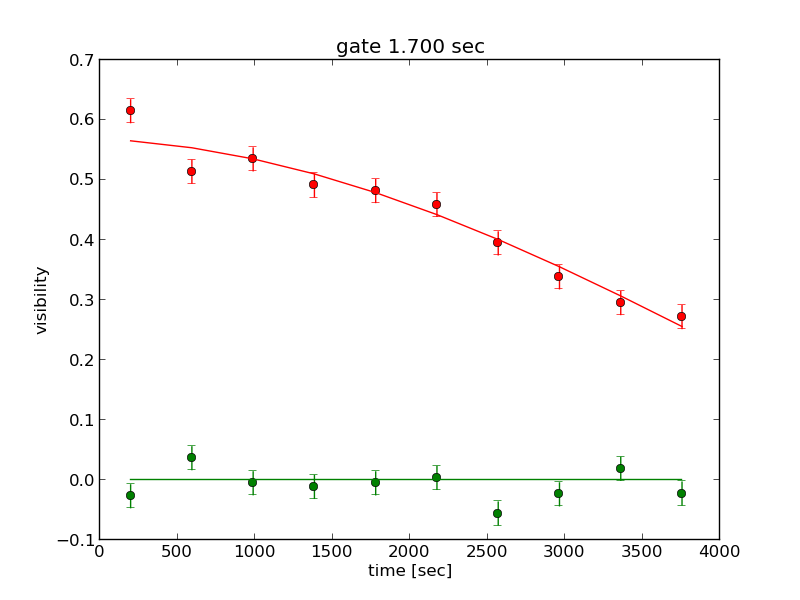}

\onefig{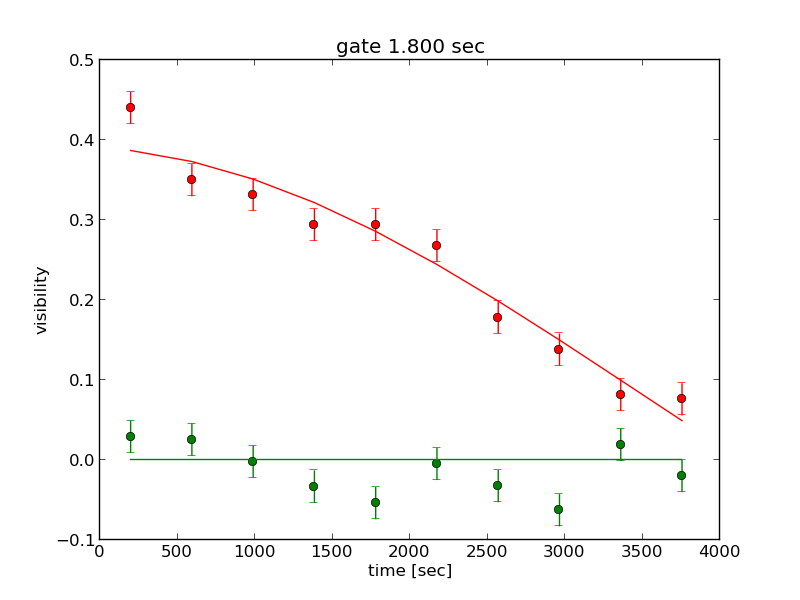}\hspace{0.5em}\onefig{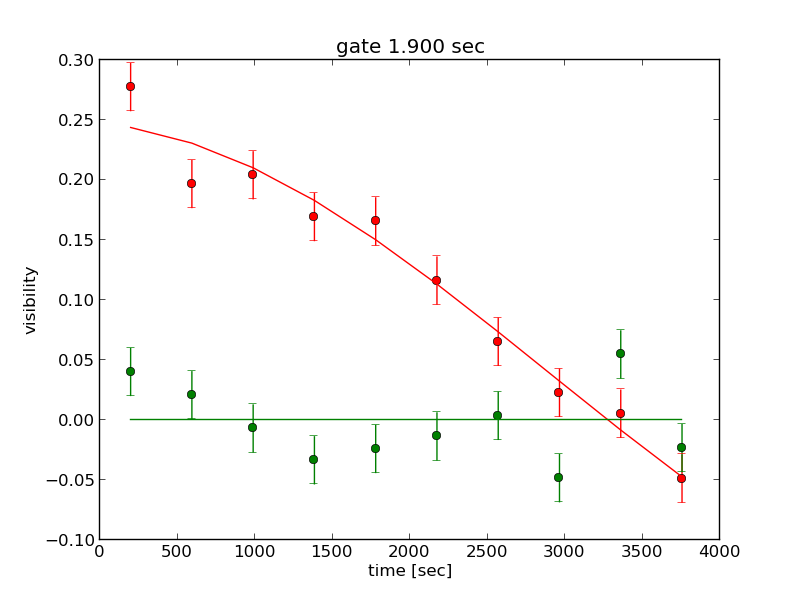}\hspace{0.5em}\onefig{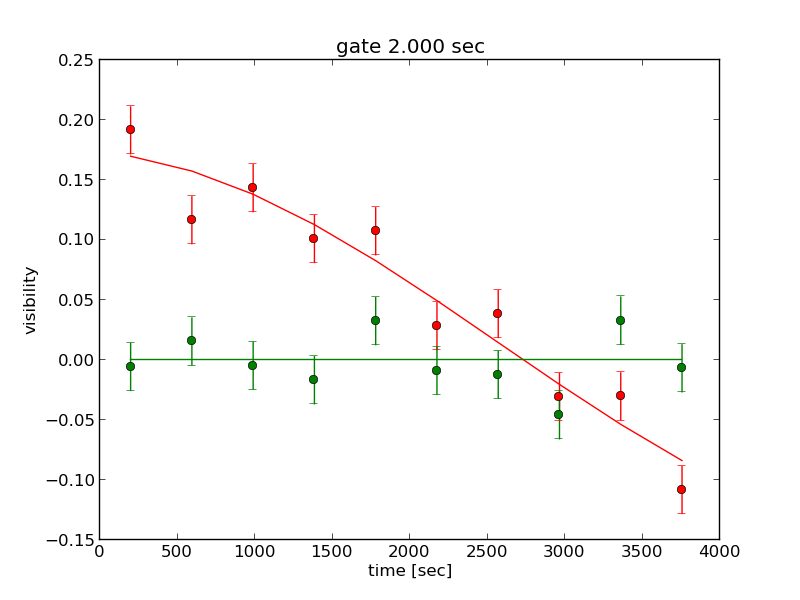}\hspace{0.5em}\onefig{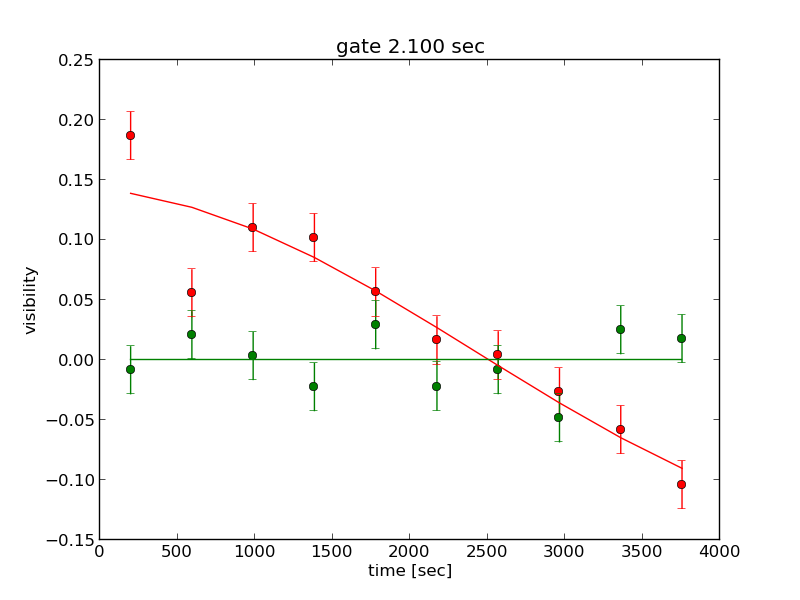}
\caption{Visibility functions for the same bins as in
  Fig.~\protect\ref{fig:visib tau}. The real part is shown in red, the
  imaginary part in green. Error bars are measurements, curves are
  model fits of narrow rings. There are still systematic deviations
  resulting from flux variability, but the main structure is reproduced
  well.}
\label{fig:visib fit rings}
\end{figure}

As Fig.~\ref{fig:visib tau} shows, apparently the size of the
scattering disk does indeed change with scattering delay, which proves
that both effects are at least partly due to the same screen. The
expectation is that, as function of delay, we see a circular ring that
expands as $\theta\propto\sqrt\tau$. To test this, we fitted (again in
$uv$-space) models of uniform (this will be refined in the future)
circular rings with arbitrary radius and total flux to the
data. Fig.~\ref{fig:visib fit rings} shows the model visibilities
compared to the measurements.

\begin{figure}[hbt]
\centering\includegraphics[angle=-90,width=0.5\textwidth]{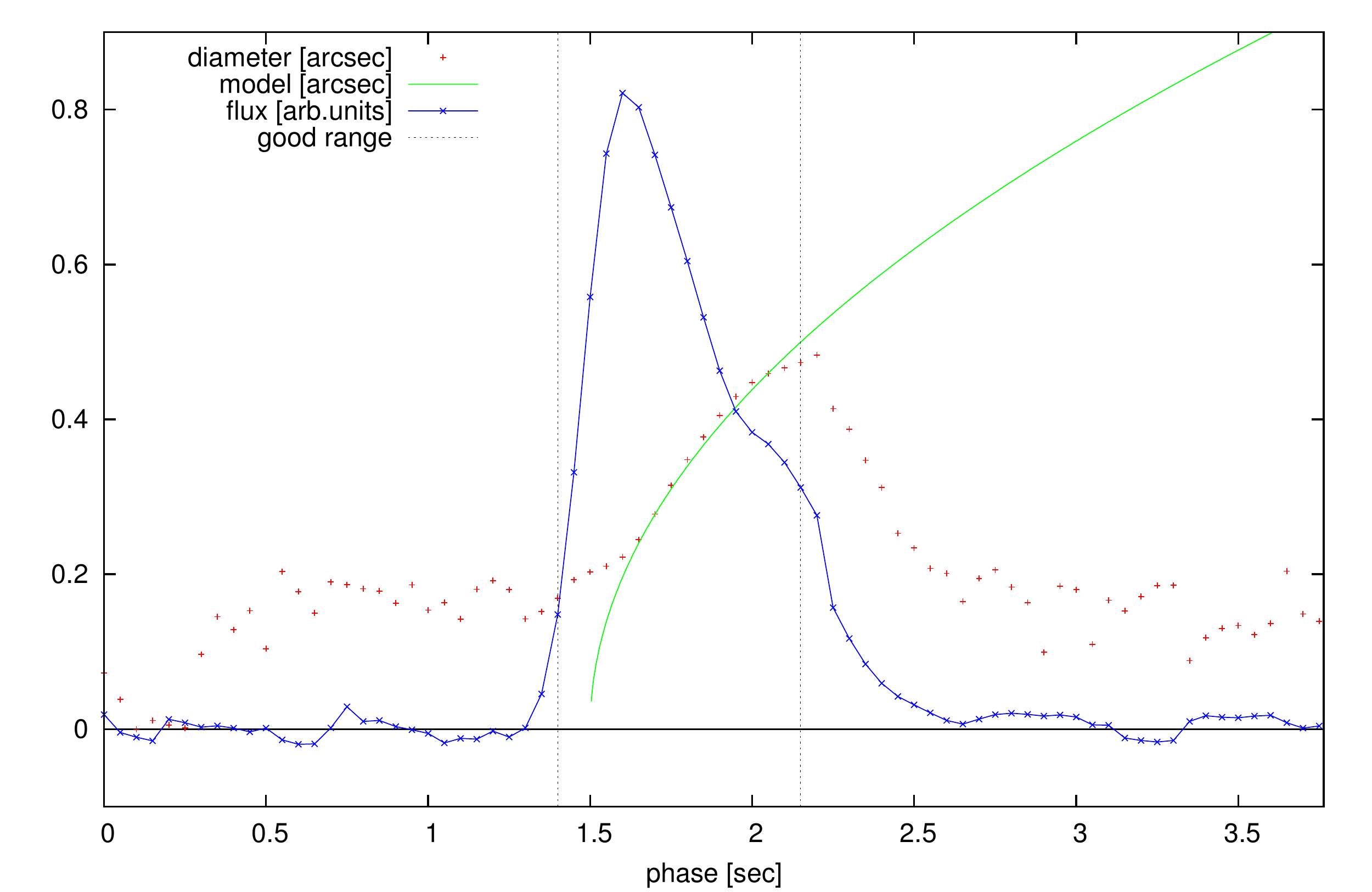}
\caption{Results from fitting the diameter of the `scattering ring' as
  function of pulse phase and thus scattering delay. The blue curve
  shows the total flux (pulse profile) and the red points are the
  diameters of the fitted rings in arcsec. Two vertical lines bracket
  the range with sufficient SNR. The green curve is a fit-by-eye of a
  $\theta\propto\sqrt\tau$ model to the most trustworthy data
  points. The deviating points near a phase of 1.5\,sec are 
  affected by the (in this preliminary analysis) low resolution in
  time and by the intrinsic pulse width.}
\label{fig:fits}
\end{figure}

Fig.~\ref{fig:fits} shows how the size of the scattering ring grows
with time. Within the best range of the pulse peak, the size seems to
follow the expected relation,
which finally confirms that for this pulsar most of the temporal and angular
scatter-broadening must be produced by \emph{one} relatively thin
screen. From the scaling constant we can estimate the effective
distance $D'=8.6\,$kpc and from that the distance of the scattering
screen from the pulsar, $\Delta=4.2\,$kpc, which localises the screen
almost exactly halfway between the pulsar (assuming it is at the same
distance as \sgra) and us.

We emphasise that this result is preliminary and that a number of
refinements have to be included in the analysis (higher
time-resolution, consider anisotropy, do consistent global fit,
consider pulsar variability, include other baselines), before the
final result can be derived and a realistic error estimate can be
attempted. For the moment our result appears to be consistent with
that of \citet{spitler14} and \citet{bower14}, but we have now
actually tested and proven the basic single thin screen assumption on
which their analysis was based.

In view of this, we have to re-consider the models of the GC
\citep[e.g.][]{lazio98}, taking into account the observational facts
that went into those models, but also our new results and the
arguments presented by \citet{langevelde92}. Finding the strongest
scattering in a small region of the sky near \sgra\ is a strange
coincidence, if the material causing it is not actually located
directly around the GC, but apparently this conclusion cannot be
avoided anymore.

Recent observations at higher frequencies (Spitler et al., in prep.)
seem to indicate strong variations in the scattering delay with
time. The most plausible explanation would be additional fast moving
clouds near the GC. If they are sufficiently small, they would not
affect the scattering size at our frequencies, and indeed they have to
be small to explain the rapid variations. In the final analysis of our
data (and of new observations in Nov 2014) we will have to take into
account possible additional scattering material closer to the pulsar,
in order to finally derive a picture that explains all the
observational data consistently. This will form invaluable input for
future pulsar searches near the black hole that is believed to be
located in the centre of our Galaxy.

\section{Acknowledgements}

The author thanks the LEAP team for providing 
observing time and for help with the observations and the data
transfer. Ralph Eatough is thanked for interesting discussions about
the subject.

{\small\bibsep0.ex
\bibliographystyle{aa}
\bibliography{proc}}

\end{document}